\documentclass{pasj00}
\twocolumn

\begin{document}
\SetRunningHead{Wang et al.}{Circular polarimetry of 4U 0142+61}
\Received{2011 Nov. 22}%{yyyy/mm/dd}
\Accepted{2012 Jan. 21}%{yyyy/mm/dd}
%\Published{}%{yyyy/mm/dd}

\title{Subaru constraint on circular polarization in I-band emission from the Magnetar 4U 0142$+$61\thanks{Based on data collected at Subaru Telescope, which is operated by the National Astronomical Observatory of Japan.}}

%%% begin:list of authors
% Do NOT capitalize all letters in "textsc".
%\author{Zhongxiang \textsc{Wang} %
%  \thanks{Example: Present Address is xxxxxxxxxx}}
%\affil{A-Address of Institute}
%\email{aaaaa@xxx.xxx.xx.xx}

%\author{B-Firstname \textsc{B-Familyname}}
%\affil{B-Address of Institute}\email{bbbbb@xxx.xxx.xx.xx}
%\and
%\author{C-Firstname {\sc C-Familyname}}
%\affil{C-Address of Institute}\email{ccccc@xxx.xxx.xx.xx}
%%% end:list of authors

%%% Please use the following style in case that sorting by 
%%% affiliation is impossible. 

 \author{
   Zhongxiang \textsc{Wang}\altaffilmark{1},
   Yasuyuki T. \textsc{Tanaka}\altaffilmark{2}
   and
   Jing \textsc{Zhong}\altaffilmark{1,3}}
\altaffiltext{1}{Shanghai Astronomical Observatory, 
Chinese Academy of Sciences, 80 Nandan Road, Shanghai 200030, China}
\email{wangzx@shao.ac.cn}
\altaffiltext{2}{High Energy Astrophysics group,
Institute of Space and Astronautical Science,
Japan Aerospace Exploration Agency,
3-1-1 Yoshinodai, Chuo-ku, Sagamihara, Kanagawa 252-5210, JAPAN}
\email{tanaka@astro.isas.jaxa.jp}
\altaffiltext{3}{Graduate School of Chinese Academy of Sciences,
No. 19A, Yuquan Road, Beijing 100049, China}
\email{jzhong@shao.ac.cn}

%% `\KeyWords{}' always has to be placed before `\maketitle'.
 %Do NOT move this preamble from here!

\maketitle

\begin{abstract}
We present the first imaging circular polarimetry of the anomalous X-ray
pulsar (AXP) 4U~0142+61 at optical wavelengths. The AXP is the only magnetar 
that has been well studied at optical and
infrared wavelengths and is known to have a complicated broad-band spectrum
over the wavelength range.  
The optical polarimetric observation was carried out with 
the 8.2-m Subaru telescope at I-band.
From the observation, the degree of circular polarization $V$
was measured to be $V=1.1\pm$2.0\%, or $|V|\leq$4.3\% (90\% confidence). 
The relatively large uncertainty was due to
the faintness of the source ($I=23.4$--24.0). 
Considering the currently suggested models 
for optical emission from magnetars, our result 
is not sufficiently conclusive to discriminate the models.
We suggest that because linear polarization is expected to be strong in 
the models, linear polarimetry of this magnetar should be conducted.
\end{abstract}

\section{Introduction}

Supported by extensive observational studies over the past 10 years, 
it is generally believed that anomalous X-ray pulsars (AXPs) and soft Gamma-ray
repeaters are magnetars---young neutron stars possessing 
ultra-high magnetic fields of $\gtrsim$10$^{14}$~G (for reviews see, 
e.g., \cite{wt06,mer08}). The magnetars are remarkable, exhibiting
a variety of high-energy phenomena \citep{kas07,mer08},
and thus have attracted great attention after their magnetar nature
was realized \citep{td96}.  While magnetars are classified as high-energy 
X-ray sources, it has been learned that they also have relatively 
strong optical and near-infrared (NIR) emission, and are detectable 
at the wavelengths as long as they are either close with 
low extinction or in bright states (i.e., in X-ray outbursts or 
flares; \cite{kas07, mer11}). 

\begin{figure}
  \begin{center}
    \FigureFile(80mm,80mm){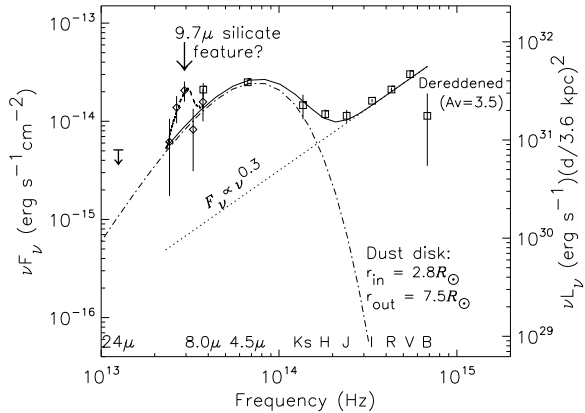}
    %%% \FigureFile(width,height){filename}
  \end{center}
  \caption{Optical and IR broad-band and \textit{Spitzer} IRS spectrum 
of the AXP 0142+61 \citep{wck08}. The squares are the optical, near-IR, 
\textit{Spitzer}\ IRAC 4.5/8.0 $\mu$m 
broadband fluxes (dereddened with $A_V=3.5$ mag), showing that the optical 
spectrum is consistent with being a power law (dotted line), 
$F_{\nu}\propto\nu^{0.3}$, and the IR spectrum can be fit with an X-ray 
irradiated dust disk model (dash-dotted curve). The diamonds are 
the dereddened IRS flux measurements, which appear as a bump when compared to
the dust disk model SED and can be fit with a silicate emission feature 
(dashed curve; \cite{slo+03}).
The MIPS 24 $\mu$m upper limit is also included in the figure.}
\label{fig:sed}
\end{figure}

Among over 20 known magnetars (McGill AXP online catalog\footnote{www.physics.mcgill.ca/pulsar/magnetar/main.html}), 
the AXP 4U 0142+61 stands out as the best studied magnetar at optical 
and IR wavelengths due to its relatively short distance 
(distance $d\simeq 3.6$ kpc) and low extinction ($A_V\simeq3.5$; 
\cite{dv06a,dv06b}).  It was the first magnetar discovered with an optical 
counterpart \citep{hvk00}, and its optical emission was found to be pulsed 
at its spin period with a pulsed fraction of 27\%
\citep{km02}, actually higher than that in its X-ray emission
(4\%--14\%; \cite{gon+10}). Aiming to search for supernova fallback disks
around young, isolated neutron stars, \citet{wck06} discovered mid-infrared
(MIR) emission from this magnetar with \textit{Spitzer Space Telescope} 
observations,
and they showed that its optical and IR spectral energy distribution (SED) 
can be described by a two-component model: one a power-law spectrum over 
optical $V\/RI$ and NIR $J$ bands probably arising from the magnetosphere
of the pulsar, and one thermal blackbody-like over the 2.2--8 $\mu$m range
arising from a debris disk. Results from follow-up \textit{Spitzer} MIR 
spectroscopy and 24~$\mu$m imaging of the source were consistent with 
their model (see Figure~\ref{fig:sed}; \cite{wck08}).
The two-component model remains controversial, as \citet{dv06c} found
that the source's NIR $K$-band flux can be highly variable with no 
correlated variability seen in its X-ray emission. In addition 
optical flux variations were also found by \citet{dv06c}, 
although not as pronounced as that in $K$-band. 

In order to fully understand optical and IR emission from magnetars,
in this paper we report our observational study of another aspect of 
4U~0142+61: circular polarization of I-band optical emission from the source.
The AXP was observed a few times in 1994--2003, and found to 
have $I=23.4$--24.0 \citep{dv06c}.

\section{Observation and Data Reduction}

Circular imaging polarimetry of 4U 0142+61 at I-band was carried out with
the 8.2-m Subaru Telescope on 2009 October 24. The imaging instrument
used was the Faint Object Camera and Spectrograph (FOCAS; \cite{kas+02}), which
can perform polarimetry with a Wollaston prism and a quarter-wave retarder
inserted to the collimated beam. The Wollaston prism splits an incident beam
into two orthogonally polarized beams, one ordinary (o-beam) 
and the other extraordinary (e-beam).  The quarter-wave plate converts 
circular polarized light into linear polarized light by retarding one 
of the beams by 1/4 of a wave.  A standard mask, provided with the FOCAS 
for imaging polarimetry to avoid blending of the two beams, was used. 
The detector was two fully-depleted-type 2k$\times$4k CCDs, with a pixel 
scale of 0.104\arcsec/pixel. The detector was 2$\times$2 binned in 
our observation.

Multiple sets of five 5-min exposures of the target field were taken with 
the quarter-wave plate at position angle (PA) 18\arcdeg\ and 108\arcdeg\ 
alternately.  Between the sets of the exposures, the telescope was dithered 
to avoid bad pixels on the CCDs. In total, we took 25 PA=18\arcdeg\ and 
20 PA=108\arcdeg\ exposures. The observing conditions were good, with 
the seeing (FWHM of point sources) varying between 0.4\arcsec--0.8\arcsec. 

\begin{figure}
  \begin{center}
    \FigureFile(80mm,80mm){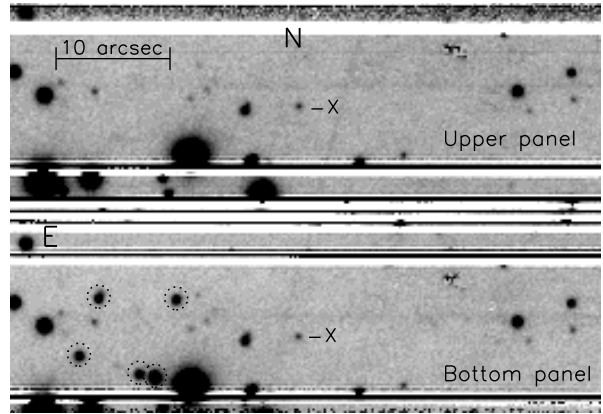}
    %%% \FigureFile(width,height){filename}
  \end{center}
  \caption{Subaru circular polarization image of the AXP 4U~0142+61 at I-band.
The same source field from the two polarized beams was recorded at the
upper and bottom panels in the image. The counterpart to 
the AXP is indicated
by $X$. Several ghost stars, marked by dotted circles, are present in 
the bottom panel.}
\label{fig:fov}
\end{figure}

We used the IRAF packages for data reduction. The images were bias subtracted
and flat fielded. Dome flats at the two position angles were taken and used
for flat fielding respectively. The images made at each PA were then 
positionally calibrated to a reference image that has the best quality, 
and were combined into one final image of the target field. 
For the sets of images at PA=18\arcdeg, a few had the seeing
larger than $\sim0.65\arcsec$ and were excluded from combining. 
In our observation, a few ghost sources appeared (see Figure~\ref{fig:fov}).
Such ghost sources are produced by the polarizer when
a source field contains bright and saturated stars. 
In our case, two ghost sources were present close to the target in three images 
of each PA sets. To filter out the two ghost sources, pixels at each pixel 
position larger than the median of the images 
by 3 standard deviations were rejected in combining. In addition, for 
the PA=108\arcdeg\ sets, one image with the two ghost sources was not included
in combining in order to cleanly remove them.
In Figure~\ref{fig:fov}, a combined image of the source field is shown.
The resulting FWHM of point sources in the two combined images is 
approximately 0.5\arcsec.  The total on-source times
are 100 min and 95 min for the images at PA=18\arcdeg\ and 
PA=108\arcdeg, respectively. 

We used {\tt DOPHOT} \citep{sms93}, a point-spread function (PSF) 
fitting photometry program, to measure brightnesses of our target and 
other in-field stars.  The obtained instrumental magnitudes and uncertainties 
are summarized in Table~\ref{tab:phot}.
In addition, aperture corrections were also applied. 
A large radius of 7.0 pixels (1.46\arcsec), 
which well included all photon counts of a point source, 
was used for photometry, and
the aperture corrections were derived using 5\ in-field, relatively 
bright stars (bright stars in the field were saturated).
The uncertainties on the corrections
were 0.021--0.035 mag, smaller 
than those from photometry (0.039--0.048 mag; Table~\ref{tab:phot}). 

We performed aperture photometry as an additional check on our results. 
In order to minimize uncertainties, a small aperture 
radius of 2.5 pixels (0.52\arcsec) was used. Aperture corrections to
a radius of 7.0 pixels were also applied.
We obtained nearly the same brightness measurements, only with uncertainties
generally 0.01~mag larger than those obtained from PSF fitting.
\begin{table}
\caption{PSF fitting photometry of 4U~0142+61}
\label{tab:phot}
  \begin{center}
    \begin{tabular}{lccc}
\hline
Sub-image & $m_{\rm pf}$ & $\Delta m_{{\rm cor.}}$ & $m_{r=7}$ \\ 
	  &  (mag)          &   (mag)     & (mag)           \\\hline
e-beam$_{18\arcdeg}$ & 24.016$\pm$0.039 & 0.326$\pm$0.021 & 23.69$\pm$0.04 \\
o-beam$_{18\arcdeg}$ &	23.977$\pm$0.040 & 0.318$\pm$0.026 & 23.66$\pm$0.05 \\
e-beam$_{108\arcdeg}$ &23.923$\pm$0.049 & 0.352$\pm$0.026 & 23.57$\pm$0.06 \\
o-beam$_{108\arcdeg}$ & 23.933$\pm$0.048 & 0.330$\pm$0.035  & 23.60$\pm$0.06 \\ 
\hline
    \end{tabular}
\vskip 1mm
\footnotesize{Note: instrumental magnitude $m=25-2.5\log({\rm flux})$; 
$m_{\rm pf}$, $\Delta m_{{\rm cor.}}$, and $m_{r=7}$ are magnitudes 
obtained from PSF fitting, aperture corrections  
to a radius of 7 pixels, and corrected magnitudes.}
%%\footnote{slsljs}
  \end{center}
\end{table}

%%%%%%%%%%%%%%%%%%%%%%%%%%%%%%%%%%%%%%%

\section{Results}

The Stokes $V$ parameter measures the degree of circular polarization, which 
is derived from
\[
V=\frac{R_V-1}{R_V+1}\ ,
\]
where
\[
R^2_V=\frac{(I_e/I_o)_{18\arcdeg}} {(I_e/I_o)_{108\arcdeg}}\ .
\]
Here $I_e$ and $I_o$ are intensities of the target in the e-beam and o-beam
frames, respectively. 
Using the formula and intensity values obtained from PSF-fitting photometry
(Table~\ref{tab:phot}), we found $V=1.1$\%, with an uncertainty of
2.0\%. 
If the aperture corrected intensities are used, the result is nearly the same
but with a slightly larger uncertainty, $V=1.4\pm2.4$\%.
Therefore from our observation, we found a 90\%-confidence constraint
of $|V|\leq 4.3$\% on 
the degree of circular polarization at I-band for 4U~0142+61. 

\section{Discussion}

Using the 8.2-m Subaru telescope, we have for the first time observationally 
studied circular polarization in optical emission from the AXP 4U 0142+61. 
We found that in the source's I-band emission, Stokes $V$ parameter 
was consistent 
with being zero, although the uncertainty was relatively large.
We note that the interstellar medium can produce polarized light by 
scattering from aligned, elongated 
dust grains, but the degree of interstellar circular polarization is 
on the order of 10$^{-4}$ (see, e.g., \cite{ave+75}), negligible to our case.
Considering that the polarimetric observation of the AXP was over 3.3 hours 
under the good-seeing conditions, 
it will be difficult to improve our measurement significantly due to the
faintness of the source. 

Currently, the origin of optical emission from magnetars is not clear. 
\citet{egl02} have suggested that their optical emission could be due to
synchrotron radiation from electron/positron pairs in ultra-high 
$B\sim 10^{15}$~G fields, similar to radio emission from radio pulsars while
scaled-up to optical wavelengths. If this is the case, 
since pulsars' radio emission is seen to be circularly 
polarized with the degree of polarization (average values over pulse profiles)
in a wide range from a few percent to as high as 
60\% (e.g., \cite{han+98,gl98}), similar polarization would be expected 
in optical emission from magnetars. Indeed,
\citet{egl02} specifically discussed the polarization detection as a method
to verify their model.  Our measurement suggests zero circular polarization
in I-band emission from 4U~0142+61, not supporting their model. However 
given the relatively large uncertainty, it is not sufficiently conclusive.
We note that the degree of linear polarization of pulsars' radio emission 
is generally 
much stronger than that of circular polarization. Even at optical wavelengths,
5 pulsars were studied through linear polarimetry and found to have 5-10\%
degree of polarization (\cite{slo+09} and references therein). 
Linear polarimetry of 4U~0142+61 should thus be conducted. Such an 
observation can be challenging. In
order to conduct linear polarimetry as deep as our circular polarimetry,
since imaging at four PAs is required, a significant amount of telescope 
time with stable, excellent seeing conditions will be needed.

In their hot corona model around magnetars, \citet{bt07} have suggested
two possible mechanisms for optical emission from magnetars: ion cyclotron 
emission or curvature emission by electron/positron pairs. In the first 
mechanism, ions in the corona of a magnetar absorb radio and microwave
radiation at their cyclotron resonance and re-emit radiation at optical/IR
wavelengths. If this mechanism is responsible for optical emission, 
certain degree of circular polarization might be present, depending on
our viewing angle of cyclotron radiation. However it is difficult
to estimate the polarization degree, thus not allowing to compare it with our 
derived upper limit. In the second mechanism, strong linear polarization 
is expected. 

As a summary, we present the first optical circular polarimetry of the AXP
4U~0142+61, the only magnetar that has been well studied at optical and IR 
wavelengths and is known to have a complicated optical and IR 
broad-band spectrum. From our observation, a 90\%-confidence constraint 
of $|V|\leq 4.3$\% in its
I-band emission is obtained. Considering the current models proposed to explain
optical emission from magnetars, the upper limit is not sufficiently conclusive
to discriminate the models. As strong linear polarization in optical emission
is expected in the models, deep optical linear polarimetry should be
conducted.

\bigskip

We thank the referee for valuable suggestions.
This research was supported by the starting funds of Shanghai
Astronomical Observatory and National Basic Research Program of China
(973 Project 2009CB824800). ZW is a Research Fellow of the 
One-Hundred-Talents project of Chinese Academy of Sciences.

%%%
% See the manual for the detail.
%%%

\end{document}